\documentclass[ prl, twocolumn, superscriptaddress, amsfonts, amsmath,floatfix]{revtex4-1}
\pdfoutput=1
\usepackage{graphicx}
\usepackage{amssymb,amsmath}
\usepackage{xspace}
\usepackage[usenames, dvipsnames]{color}
\usepackage{natbib}

\newcommand{\be}{\begin{equation}}
\newcommand{\ee}{\end{equation}}

\newcommand{\rg}{{\bf r}}
\newcommand{\Eg}{{\bf E}}
\newcommand{\Dg}{{\bf D}}

\newcommand{\dt}{\langle \Delta t \rangle}

\begin{document}

\title{Causality and instability in wave propagation in random time-varying media}
\author{R. Pierrat}
\altaffiliation{romain.pierrat@espci.psl.eu}
\affiliation{Institut Langevin, ESPCI Paris, PSL University, CNRS, 1 rue Jussieu, 75005 Paris, France}
\author{J. Rocha}
\affiliation{Institut Langevin, ESPCI Paris, PSL University, CNRS, 1 rue Jussieu, 75005 Paris, France}
\author{R. Carminati} 
\affiliation{Institut Langevin, ESPCI Paris, PSL University, CNRS, 1 rue Jussieu, 75005 Paris, France}
\affiliation{Institut d'Optique Graduate School, Paris-Saclay University, 91127 Palaiseau, France}

\begin{abstract}
   We develop a theoretical model to investigate wave propagation in media with random time-varying properties, where
   temporal fluctuations lead to complex scattering dynamics. Focusing on the ensemble-averaged field, we derive an
   exact expression for the average Green's function in the presence of finite temporal disorder, and extend the
   analysis to the thermodynamic limit. In contrast to spatial disorder, causality prevents recurrent scattering,
   allowing us to achieve a non-perturbative solution. We introduce an effective medium description providing a simple
   analysis of the propagation regimes. Our findings offer new insights into wave dynamics in temporally disordered
   media, with potential applications in time-varying metamaterials, dynamic sensing, and imaging in turbulent or
   chaotic environments.
\end{abstract}

\maketitle

Wave propagation in random time-varying media, where the medium properties fluctuate randomly over time, presents a
challenging and intriguing problem. Unlike traditional settings involving spatially disordered static media, this
scenario introduces randomness in the temporal domain, resulting in complex interactions between waves and the
time-dependent fluctuations of the medium. Such environments are relevant in numerous physical systems where the wave
velocity or refractive index changes randomly over time, ranging from electromagnetic waves in fluctuating
plasmas~\cite{KalluriBook} or wireless information networks~\cite{Broadwater1993} to acoustic waves in temporally
varying atmospheres~\cite{Kallistratova2002}. Recently, the subject has expanded considerably with the emergence of
temporally controllable metamaterials in acoustics~\cite{Zangeneh2019}, electromagnetics~\cite{Caloz2020} and
optics~\cite{Galiffi2022} and with the concept of temporal interface as a building block~\cite{Mostafa2024}. When the
modulation of the material properties is periodic in time, bandgaps appear in the wave propagation constant
$k$~\cite{Holberg1966,Zurita2009}, as well as topological phases~\cite{Lustig2018}, leading to the concept of photonic
time crystal~\cite{Saha2023}. Other types of deterministic modulation have revealed rich analogies between temporal and
spatial structuring of the propagation medium~\cite{Bacot2016,Pacheco2020,Tirole2023}. 

Random fluctuations in time give rise to unique wave phenomena, including multiple scattering effects that are not
present in spatially disordered systems. For example, it has been shown that a pulse subjected to a series of random
kicks, corresponding to sudden changes in the wave velocity (a model very close to that used in this work), has an
ensemble-averaged energy increasing exponentially with time at long
times~\cite{Sharabi2021,Carminati2021,Garnier2021,Apffel2022}. In this regime, the statistical distribution of the
energy becomes lognormal~\cite{Carminati2021}. The combination of these two features reveals an analogy with Anderson
localization in spatially disordered one-dimensional systems~\cite{Anderson1980}. While many studies have focused on the
transport of wave energy, it seems that the statistical behavior of the field itself has been overlooked. Here we
specifically investigate the ensemble-averaged field, providing new insights into how random fluctuations and causality
affect a wave undergoing multiple scattering in temporal disorder.

In this Letter, we develop a theoretical model to calculate the average field in a medium with random time-varying
properties. Our analysis includes the case of a finite-duration temporal disorder, where we demonstrate that the problem
is exactly solvable. In the limit of an infinite temporal disorder, we find that the independent scattering
approximation becomes exact due to the constraints imposed by causality, putting forward a major difference with wave
transport in pure spatial disorder. In addition, we show that the average field is characterized by an effective
frequency that depends on the average value of the random time fluctuations. We also identify a peculiar regime in which
the average field grows exponentially at long times. The behavior of the average field versus the average energy for
this problem is analyzed in terms of the stability of a noisy parametric oscillator.  This work provides a rigorous
framework for understanding wave propagation in random time-varying media, offering new perspectives on the interplay
between temporal randomness and wave dynamics, with potential implications for wave-based technologies where temporal
fluctuations play a crucial role, such as time-varying metamaterials, dynamic sensing and imaging, and communication
networks.

A scalar wave with amplitude $\psi(\rg,t)$ propagating in a homogeneous medium with a time-varying velocity obeys the equation
\begin{equation}
   \nabla^2 \psi(\rg,t) - \frac{\varepsilon(t)}{c^2}\frac{\partial^2 \psi}{\partial t^2}(\rg,t) = 0 \, ,
   \label{eq:wave_1D}
\end{equation}
where $\varepsilon(t)$ is a real-valued time-modulated function, and $c$ is the wave velocity in the unmodulated
background medium. This model applies to different kinds of waves, including acoustic waves, surface waves on liquids,
or electromagnetic waves~\cite{comment1}, in the absence of dispersion and intrinsic damping. The effect of dispersion
in time-varying media has been introduced and discussed recently~\cite{Sotoodehfar2022,Koutserimpas2024}, and is left
for future developments of the present theory.

For a field depending on a spatial coordinate $x$, the Green's function connecting a spacetime souce point $(x',t')$ to
a spacetime observation point $(x,t)$ is of the form $G(x-x',t,t')$. In $k$-space, the Green's function $G(k,t,t') =
\int_{-\infty}^{+\infty} G(X,t,t')  \exp(-ikX) dX$ is defined as the causal solution to 
\begin{equation}
   \left [ \frac{\partial^2}{\partial t^2} + \Omega^2(t) \right ] G(k,t,t') = \delta(t-t') \, ,
   \label{eq:Green_general}
\end{equation}
where $\Omega^2(t) = c^2 k^2/\varepsilon(t)$. We point out that the model is not limited to a homogeneous space, and
could be adapted, for example, to a waveguide filled with a homogeneous time-dependent medium with $k$ the wavevector of
a guided mode. In this work, we consider random fluctuations of the medium, and $\Omega^2(t)$ is a realization of a
random process. In practice, it seems relevant to consider that $\varepsilon(t)$ keeps positive values, and we will
assume $\Omega^2(t) \geq 0$. We are interested in the behavior of the average Green's function $\langle G(k,t,t')
\rangle$, where the brackets denote the ensemble average over a set of realizations of $\Omega^2(t)$.

In the absence of time-modulation, $\Omega^2(t) = c^2k^2 \equiv \Omega^2_b$ is a constant function, and the Green's
function $G_b$ in the background medium is the causal solution to Eq.~(\ref{eq:Green_general}) with $\Omega^2(t)$
replaced by $\Omega^2_b$. Its expression is $G_b(k,t,t')=\operatorname{H}(\tau)\sin(\Omega_b \tau)/\Omega_b$, where
$\tau=t-t'$ and $\operatorname{H}(\tau)$ is the Heaviside step function~\cite{Selvestrel2024}. Subtracting the equations
satisfied by $G$ and $G_b$, we find that the scattered Green's function $G_s=G-G_b$ also satisfies
Eq.~(\ref{eq:Green_general}) with $\Omega^2(t)$ replaced by $\Omega^2_b$, and the delta function in the right-hand side
replaced by the source term $V(t) G(k,t,t')$, where $V(t) = \Omega^2_b - \Omega^2(t)$ is the scattering potential. We
deduce that $G_s(k,t,t') = \int G_b(k,t,t'') V(t'') G(k,t'',t') \, dt''$ which, in terms of the full Green's function
$G$, is also
\begin{equation}
   G(t,t') = G_b(t,t') + \int G_b(t,t'') V(t'') G(t'',t') \, dt'' \, .
   \label{eq:integral_G}
\end{equation}
This self-consistent integral equation is a usual starting point in multiple scattering
theory~\cite{ShengBook,AkkermansBook,CarminatiBook}. For simplicity we drop, from now on, the variable $k$ in the
Green's functions, keeping in mind that all calculations are performed at a fixed spatial frequency $k$. 

\begin{figure}[h]
     \begin{center}
        \includegraphics[width=8cm]{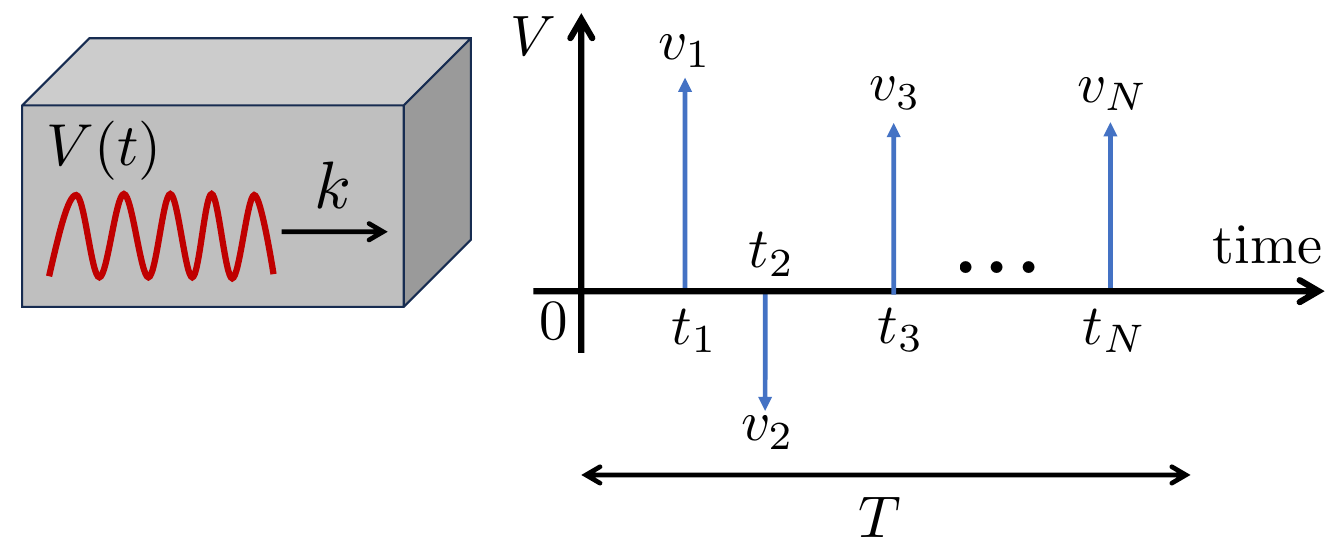}
     \end{center}    
     \caption{\label{fig:delta_kicks} Left: Wave at a fixed spatial frequency $k$ propagating in a homogeneous medium
     with a random potential $V(t)$. Right: Random chain of kicks. The kick strenghts $v_j$ and the times $t_j$ are
     random variables.}
\end{figure}

To proceed further, we model the time-disordered medium with a chain of $N$ infinitely short kicks, randomly distributed
over a time window $T$, as represented schematically in Fig.~\ref{fig:delta_kicks}. 
We write the potential in the form $V(t) = \sum_{j=1}^{N} v_j \delta (t-t_j)$, where the positions $t_j$ of the kicks
and their amplitudes $v_j$ (assumed to be real valued and finite) are statistically independent random variables, which
is a simple model for an uncorrelated temporal disorder~\cite{comment}. Note that in this setting negative values of the
potential $V(t)$ are allowed, corresponding to $\Omega^2(t) > \Omega_b^2$. The delta functions represent kicks with
duration much smaller than the intrinsic time scale of the incident wave and the time between successive kicks, which is
an assumption of the present theory. We also assume that the $t_j$ are uniformly distributed over the time window
$[0,T]$. Using this expression of $V(t)$ in Eq.~(\ref{eq:integral_G}), and developing the Born series, we obtain
\begin{align}
   G(t,t') = G_b(t,t') + \sum_{j=1}^{N} G_b(t,t_j) v_j G_b(t_j,t') \nonumber \\
   + \sum_{j=1}^{N} \sum_{l=1}^{N}G_b(t,t_l) v_l G_b(t_l,t_j) v_j G_b(t_j,t') + ...
   \label{eq:Born_series}
\end{align}
Note that the causal time ordering of the scattering events is naturally accounted for by the Heaviside function in
$G_b$, and does not need to be included in the statistics of the times $t_j$. 

We will now show that in the absence of statistical correlations in the temporal disorder ({\it i.e.} all random
variables $v_j$ and $t_j$ are statistically independent), the average Green's function $\langle G(t,t') \rangle$ can be
computed explicitly, even for finite-size time disorder. Importantly, we will consider that the observation and source
times $t$ and $t'$ lie within the disorder window $[0,T]$. Averaging Eq.~(\ref{eq:Born_series}), the $(n+1)$-th term
takes the form
\begin{align}
   \sum_{\alpha_1, ... \alpha_n} \int dv_{\alpha_1} ... dv_{\alpha_n}   dt_{\alpha_1} ... dt_{\alpha_n} \, f(v_{\alpha_1},... ,v_{\alpha_n},t_{\alpha_1}, ... ,t_{\alpha_n}) \nonumber \\
   \times G_b(t,t_{\alpha_1}) v_{\alpha_1} G_b(t_{\alpha_1},t_{\alpha_2}) ... G_b(t_{\alpha_n},t') \, ,
   \label{eq:G_interm}
\end{align}
where $f$ is the probability density of the kicks amplitudes and times. According to the hypotheses above, it factorizes into
\begin{equation}
   f(v_{\alpha_1},... ,v_{\alpha_n},t_{\alpha_1}, ... ,t_{\alpha_n}) = \frac{1}{T^n} P(v_{\alpha_1}) P(v_{\alpha_2}) ... P(v_{\alpha_n}) \, ,
   \label{eq:density_factorized}
\end{equation}
with $P$ the probability density of the kick amplitudes (the calculation is independent of the shape of this probability
density). Using Eqs.~(\ref{eq:Born_series})-(\ref{eq:density_factorized}), we find that
\begin{equation}
   \langle G(t,t') \rangle = \sum_{n=0}^N \frac{N!}{(N-n)!} \frac{\langle v \rangle^n}{T^n} G_n(t,t') \, ,
   \label{eq:G_av_N}
\end{equation}
where $\langle v \rangle$ is the average kick amplitude and $G_n$ satisfies the recursive relationship
\begin{equation}
   G_n(t,t') = \int_0^T G_b(t,t'') G_{n-1}(t'',t') \, dt''
   \label{eq:G_n_recursive}
\end{equation}
with $G_0 = G_b$. The factor $N!/(N-n)!$ in Eq.~(\ref{eq:G_av_N}) results from the counting of terms in the summation
(\ref{eq:G_interm}) in which, as a consequence of causality, repeated scattering over the same scatterer (recurrent
scattering) is not possible. The calculation of the average Green's function amounts to evaluating $G_n$. Due to the
presence of the Heaviside function in $G_b$, the integral in Eq.~(\ref{eq:G_n_recursive}) can be extended to
$[-\infty,+\infty]$. The recursive relation takes the form of a convolution product that can be evaluated in the time
Fourier domain, as shown in the Supplemental Material~\cite{SM}. We find that $G_n(t,t') = \operatorname{H}(\tau)
\tau^{n+1} j_n(\Omega_b \tau) /(2^n \Omega_b^n n!)$, with $j_n$ the spherical Bessel function of order $n$. Finally, the
average Green's function for a series of $N$ uncorrelated kicks distributed over a duration $T$ is
\begin{align}
   \langle G(t,t') \rangle = \tau \, \operatorname{H}(\tau) \sum_{n=0}^N \frac{N!}{n! (N-n)!}
   \left [ \frac{\langle v \rangle \, \tau}{2 \Omega_b T} \right ]^n   j_n(\Omega_b\tau) \, .
   \label{eq:Gav_exact}
\end{align}

This expression of the average Green's function is the first important result of this Letter. Remarkably, this is the
exact solution to the problem of determining the average field in an uncorrelated temporal disorder with finite duration
$T$. In the case of one-dimensional space disorder, that has been extensively studied~\cite{MelloBook}, it seems that an
exact solution for a finite-size system is out of reach. For temporal disorder, causality excludes recurrent scattering
events, making a non-pertubative calculation of the average field possible without any restriction on the duration $T$
of the time window.

\begin{figure}[h]
     \begin{center}
        \includegraphics[width=8cm]{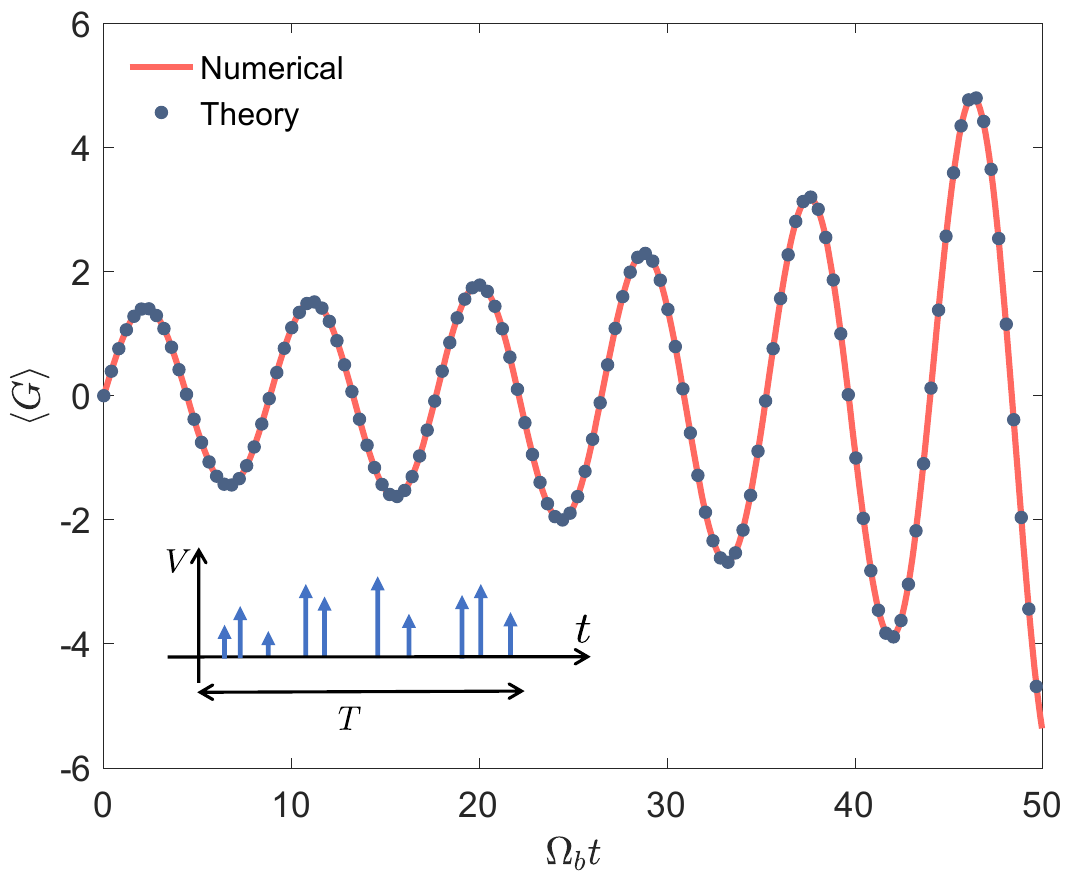}
     \end{center}    
     \caption{\label{fig:finite} Average Green's function $\langle G(t,t') \rangle$ versus time for a finite-size
     disorder. The random time modulation of the medium is activated during a finite interval $[0,T]$, with $\Omega_b T
     = 100$. The source time is $t'=0$ and the observation time $t$ lies within the interval $[0,T]$. Black solid line:
     Numerical simulation using the transfer matrix method. Markers: Direct calculation using the exact
     expression~(\ref{eq:Gav_exact}) valid for a finite-size system. As sketched in the inset, the disorder is modelled
     by a series of $N=100$ kicks, with $\langle v \rangle = 0.5 \Omega_b$ and the $v_j$ distributed uniformly in the
     interval $[0.25 \Omega_b,0.75 \Omega_b]$. }
\end{figure}

We checked the validity of expression (\ref{eq:Gav_exact}) against numerical simulations, carried out using the transfer
matrix method. The principle of the method, previously used in Ref.~\cite{Carminati2021}, is summarized in the
Supplemental Material for completeness~\cite{SM}. Given a set of random chains of kicks generated numerically, the
method allows us to compute the Green's function in each realizations, and to perform an ensemble average over the full
set. A comparison between the numerical simulation and a direct calculation from Eq.~(\ref{eq:Gav_exact}) is shown in
Fig.~\ref{fig:finite}. We observe an excellent agreement between the numerical simulation and the analytical expression~
(\ref{eq:Gav_exact}), as expected. We also see that the amplitude of the average field increases when the observation
time approaches the end $T$ of the disorder window.  Finite size effects appear to be more critical than in the case of
spatial disorder. This is due to the absence of backscattering in time, which increases the sensitivity of the field to
scattering events close to the final boundary of the disordered region, and to the fact that the average field does not
decay exponentially with time as it does in the case of spatial disorder.
%

We now examine the limit of an infinite time window with $T\to\infty$ and $N\to\infty$ at a constant kick density $N/T
\equiv 1/\dt$ (thermodynamic limit).  From Eq.~(\ref{eq:Gav_exact}), we immediately obtain
\begin{align}
   \langle G(t,t') \rangle = \tau \, \operatorname{H}(\tau) \sum_{n=0}^{+\infty} \frac{1}{n!}
   \left [ \frac{\langle v \rangle \, \tau}{2 \Omega_b \dt} \right ]^n   j_n(\Omega_b\tau) \, .
   \label{eq:Gav_thermo_interm}
\end{align}
This expression can be simplified using the identity $(1/z) \cos\sqrt{z^2-2yz} = \sum_{n=0}^{+\infty} (y^n/n!)
j_{n-1}(z)$~\cite{AbramowitzBook}, deriving with respect to $y$, and applying the resulting relationship with
$z=\Omega_b\tau$ and $y=\langle v \rangle \tau/(2\Omega_b \dt)$. This leads to
\begin{align}
   \langle G(t,t') \rangle = \operatorname{H}(\tau) \frac{\sin(\Omega_e\tau)}{\Omega_e} \ \mathrm{with} \  \Omega_e^2 =
   \Omega_b^2 - \frac{\langle v \rangle}{\dt} \, .
   \label{eq:Gav_thermo}
\end{align}
This expression is the second important result in this Letter. It shows that in an infinite time disorder (in practice,
$N \gg 1$ or equivalently $T \gg \dt$), the average Green's function takes the same form as the Green's function in the
background medium, with the frequency $\Omega_b$ replaced by an effective frequency $\Omega_e$. The effective medium
description is well-known in multiple scattering of waves in space
disorder~\cite{ShengBook,AkkermansBook,CarminatiBook}, and it may not be surprising to find a similar description with
time disorder. 

There is, however, a feature of the time domain that deserves to be highlighted. Expression (\ref{eq:Gav_thermo}), valid
in the thermodynamic limit, is also an exact (non perturbative) result. For spatial disorder, similar expressions are
found only in the independent scattering approximation (ISA), where statistical correlations in the positions of the
scatterers and recurrent scattering events are neglected~\cite{ShengBook,AkkermansBook,CarminatiBook}. In a temporal
disorder, recurrent scattering events are excluded by causality in the first place, which makes the ISA a rigorous
framework (in the absence of correlations in the disorder, which we assumed from the beginning). This feature of
propagation in time disorder can be understood from another point of view. Rewriting Eq.~(\ref{eq:G_av_N}) in the
thermodynamic limit, we find
\begin{equation}
   \langle G(t,t') \rangle = \sum_{n=0}^N \frac{\langle v \rangle^n}{\dt^n} \, G_n(t,t') \, ,
   \label{eq:G_av_N_thermo}
\end{equation}
which can be seen as an iterative solution to the Dyson equation
\begin{equation}
   \langle G(t,t') \rangle = G_b(t,t') + \int_{-\infty}^{+\infty} G_b(t,t'') \frac{\langle v \rangle}{\dt} \langle G(t'',t') \rangle \, dt'' \, .
   \label{eq:Dyson_thermo}
\end{equation}
We immediately see that the average Green's function satisfies the Dyson equation with a kernel (self-energy) $\langle
V(t) \rangle = \langle v \rangle/\dt$. This expression of the self-energy as the product of the (time) density of
scatterers $1/\dt$ and the average scattering operator $\langle v \rangle$ of a single scatterers is typical of the
ISA~\cite{ShengBook,AkkermansBook,CarminatiBook}. But for time-disorder Eq.~(\ref{eq:Dyson_thermo}) is exact, showing
again that the ISA in this setting provides rigorous results.

\begin{figure}[h]
     \begin{center}
        \includegraphics[width=8cm]{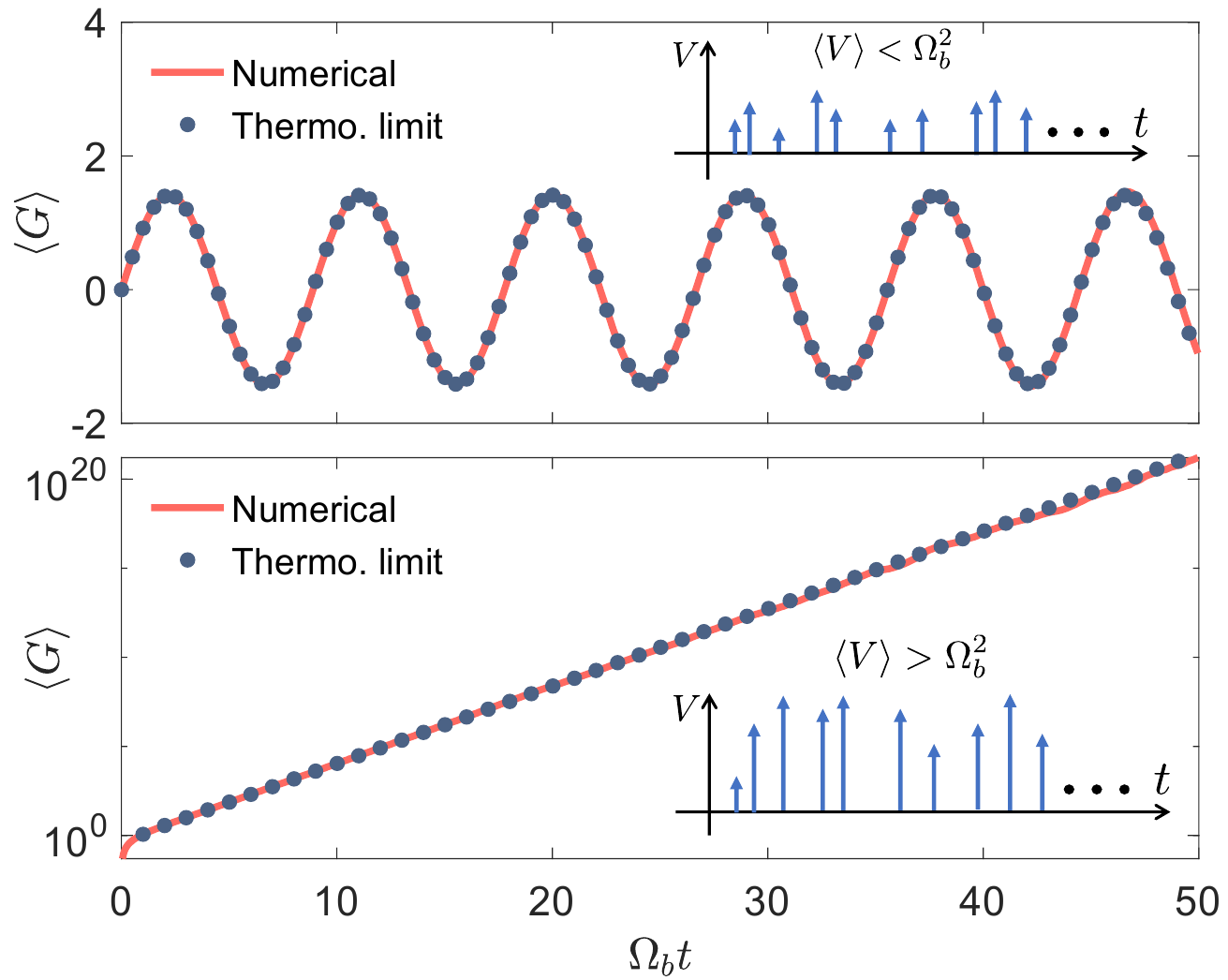}
     \end{center}    
     \caption{\label{fig:infinite} Average Green's function $\langle G(t,t') \rangle$ versus time in the thermodynamic
     limit. Comparison between numerical simulations and a direct calculation using Eq.~(\ref{eq:Gav_thermo}). The
     random time modulation of the medium is activated during a large finite interval $[0,T]$, with $\Omega_b T = 10^4$.
     The source time is $t'=0$ and for observation times satisfying $0 \ll t \ll T$ the system can be considered to be
     infinite. The disorder is modelled by a series of $N=10^4$ kicks, which corresponds to $\dt = 1/\Omega_b$. Top:
     Regime $\langle V(t) \rangle = \langle v \rangle/\dt < \Omega_b^2$, where the average Green's function oscillates
     at the effective frequency $\Omega_e$ with a constant amplitude. $\langle v \rangle = 0.5 \Omega_b$ with the $v_j$
     distributed uniformly in the interval $[0.25 \Omega_b,0.75 \Omega_b]$. Bottom:  Regime $\langle V(t) \rangle >
     \Omega_b^2$ where the average Green's function increases exponentially with time.  $\langle v \rangle = 2 \Omega_b$
     with the $v_j$ distributed uniformly in the interval $[1.75 \Omega_b,2.25 \Omega_b]$. }
\end{figure}

Having assumed $\Omega^2(t)\ge 0$ in the first place implies that $\langle V(t) \rangle =  \Omega_b^2 - \langle
\Omega^2(t)\rangle \leq \Omega_b^2$. Since $\langle V(t) \rangle = \langle v \rangle/\dt$, this also means that
$\Omega_e^2 >0$ whatever the scattering strength. The effective frequency $\Omega_e$ is real valued, and the average
Green's function oscillates in time with a constant amplitude. We also find that the effective frequency $\Omega_e$ can
be smaller or larger than the background frequency $\Omega_b$, depending on the sign of the average kick strength
$\langle v \rangle$. The behavior of the average field in the thermodynamic limit is illustrated in
Fig.~\ref{fig:infinite} (top panel), where the prediction from Eq.~(\ref{eq:Gav_thermo}) is compared to exact numerical
simulations in a sufficiently large system considered to be infinite. We see that the average Green's function
calculated numerically oscillates at an effective frequency $\Omega_e$ in agreement with the prediction from
Eq.~(\ref{eq:Gav_thermo}). 
%

It may be enlightening to examine the behavior of the average field from an alternative point of view.
Equation~(\ref{eq:Green_general}), that defines the Green's function $G$ in the presence of temporal disorder, is the
dynamic equation of a parametric oscillator with randomly driven frequency $\Omega(t)$. Random parametric oscillators
are found in many fields and have been extensively studied~\cite{Bourret1971,Bourret1973,Lindenberg1981, GittermanBook}.
For example, it is known that for a disordered potential with zero mean, the first moment of the oscillator is not
affected by randomness, which is consistent with Eq.~(\ref{eq:Gav_thermo}) in which $\langle G \rangle = G_b$ when
$\langle v \rangle = 0$. For a biased potential ($\langle v \rangle \neq 0$), we find that $\langle G \rangle$
oscillates at a shifted frequency $\Omega_e = \Omega_b \sqrt{1-\langle v \rangle / (\dt \Omega_b^2)}$.  It is also known
that in terms of average energy, which statistically corresponds to a second-order moment of the amplitude, a noisy
oscillator as described by Eq.~(\ref{eq:Green_general}) is always unstable in the absence of internal
damping~\cite{Lindenberg1981}. In our problem this corresponds to an average energy of the field growing exponentially
at long times, which is consistent with previous works~\cite{Sharabi2021,Carminati2021}.

Finally, it is interesting to note that the theory also includes the possibility of having $\langle V(t) \rangle =
\langle v \rangle/\dt > \Omega_b^2$, which corresponds to negative values of $\Omega_e^2$ in Eq.~(\ref{eq:Gav_thermo}).
This situation is peculiar since it requires $\varepsilon(t)$ to take negative values in order to reach $\langle
\Omega^2(t)\rangle < 0$, and might be relevant only from a pure theoretical standpoint. In this regime, the average
field is found to grow exponentially at long times, with a time constant $1/\operatorname{Im} \Omega_e$. The behavior of
the average field in this regime is illustrated in Fig.~\ref{fig:infinite} (bottom panel), where the prediction from
Eq.~(\ref{eq:Gav_thermo}) is compared to exact numerical simulations, with here as well excellent agreement. Physically,
a negative value of $\langle \Omega^2(t)\rangle \propto \langle 1/\varepsilon(t) \rangle$ prevents the average field to
propagate. With the modulation of $\varepsilon(t)$, the wave gets amplified without propagation, leading to a growing
stored energy in the field.

In summary, this study presents a comprehensive theoretical model to describe the average field in a medium with random
time-varying properties. Our exact solution for finite temporal disorder and its extension to the thermodynamic limit
reveal the profound role of causality in suppressing recurrent scattering events, a key distinction from wave transport
in spatial disorder. The description of the average field in terms of an effective frequency, that depends on the
average value of the random fluctuations, highlights the novel dynamics of waves in temporally disordered systems. These
findings not only provide a rigorous framework for understanding wave propagation in random time-varying media but also
have implications for various wave-based technologies, such as dynamic sensing, imaging in turbulent or chaotic
environments, and time-varying metamaterials. Our results align with numerical simulations, reinforcing the accuracy of
our analytical approach and offering new perspectives on wave dynamics in time fluctuating media. A natural extension of
the subject could include the case of correlated temporal disorder, offering an additional degree of freedom for wave
control~\cite{Kim2023}, in line with recent work on spatial disorder~\cite{Vynck2023}. 

\begin{acknowledgments}
   We thank David Gaspard, Alexandre Selvestrel and Boris Shapiro for valuable inputs. This work has received support
   under the program ``Investissements d'Avenir'' launched by the French Government.
\end{acknowledgments}

\end{document}


\title{Causality and instability in wave propagation in random time-varying media\\Supplemental Material}

\author{R. Pierrat}
\altaffiliation{romain.pierrat@espci.psl.eu}
\affiliation{Institut Langevin, ESPCI Paris, PSL University, CNRS, 1 rue Jussieu, 75005 Paris, France}
\author{J. Rocha}
\affiliation{Institut Langevin, ESPCI Paris, PSL University, CNRS, 1 rue Jussieu, 75005 Paris, France}
\author{R. Carminati}
\affiliation{Institut Langevin, ESPCI Paris, PSL University, CNRS, 1 rue Jussieu, 75005 Paris, France}
\affiliation{Institut d'Optique Graduate School, Paris-Saclay University, 92127 Palaiseau, France}

\maketitle

\section{Computation of $G_n$}

The objective of this section is to compute $G_n$ defined by
\begin{equation}
   G_{n}(t,t')=\int_{0}^{T} G_b(t,t'')G_{n-1}(t'',t')\ud t''
\end{equation}
where $G_0=G_b$. Thanks to the Heaviside function in $G_b$ and because we are considering source and observation times
lying in the range $[0,T]$ only, the bounds of the integral can be extended to $\pm\infty$. Moreover, $G_b(t,t'')$ depends only
on $t-t''$ and we define $\tilde{G}_b(t-t'')=G_b(t,t'')$ which gives
\begin{equation}
   G_{n}(t,t')=\int_{-\infty}^{+\infty} \tilde{G}_b(t-t'')G_{n-1}(t'',t')\ud t''.
\end{equation}
By recurrence, we can easily see that $G_n(t,t')$ also depends only on $t-t'$ and we define
$\tilde{G}_n(t-t')=G_n(t,t')$ which leads to
\begin{equation}
   \tilde{G}_{n}(t-t')=\int_{-\infty}^{+\infty} \tilde{G}_b(t-t'')\tilde{G}_{n-1}(t''-t')\ud t''
\end{equation}
and which is a convolution product. Applying a Fourier transform, we get
\begin{equation}
   \tilde{G}_{n}(\omega)=\tilde{G}_b(\omega)\tilde{G}_{n-1}(\omega)=\tilde{G}_b(\omega)^{n+1}.
\end{equation}
$\tilde{G}_b$ being defined as a sinus function, it is not Fourier transformable in terms of functions. To avoid
difficulties in the use of distributions, we consider a modified Green's function 
\begin{equation}
   \bar{G}_b(\tau)=\he(\tau)\frac{\sin(\Omega_b\tau)}{\Omega_b}\exp(-\alpha\tau)
\end{equation}
with $\alpha>0$. The limit $\alpha\to 0^+$ will be taken at the end of the calculation. This gives
\begin{multline}
   \bar{G}_b(\omega)=\frac{1}{\Omega_b^2+(\alpha-i\omega)^2}
\\
   \quad\Rightarrow\quad
   \bar{G}_n(\omega)=\frac{1}{\left[\Omega_b^2+(\alpha-i\omega)^2\right]^{n+1}}.
\end{multline}
We now have to compute the inverse Fourier transform
\begin{equation}
   \bar{G}_n(\tau)=\int_{-\infty}^{+\infty}\frac{\exp(-i\omega\tau)}{\left[\Omega_b^2+(\alpha-i\omega)^2\right]^{n+1}}
      \frac{\ud\omega}{2\pi}.
\end{equation}
We apply the residue method in the lower complex plane (\ie, $\im\omega<0$) since $\tau>0$. In this plane, two
poles $\omega^{\pm}=\pm\Omega_b-i\alpha$ of order $n+1$ are identified. Thus
\begin{multline}
   \bar{G}_n(\tau)=-i\he(\tau)\left[\operatorname{Res}(\omega^+)+\operatorname{Res}(\omega^-)\right]
\\
      =\frac{(-1)^{n+1}\he(\tau)}{in!}\left\{
         \lim_{\omega\to\omega^+}\frac{\ud^n}{\ud\omega^n}\left[\frac{\exp(-i\omega\tau)}{(\omega-\omega^-)^{n+1}}\right]
\right.\\\left.
        +\lim_{\omega\to\omega^-}\frac{\ud^n}{\ud\omega^n}\left[\frac{\exp(-i\omega\tau)}{(\omega-\omega^+)^{n+1}}\right]
      \right\}.
\end{multline}
Taking the limit $\alpha\to 0^+$, we find that
\begin{multline}
   \tilde{G}_n(\tau)=\frac{(-1)^{n+1}\he(\tau)}{in!}\left\{
         \lim_{\omega\to\Omega_b}\frac{\ud^n}{\ud\omega^n}\left[\frac{\exp(-i\omega\tau)}{(\omega+\Omega_b)^{n+1}}\right]
\right.\\\left.
        +\lim_{\omega\to-\Omega_b}\frac{\ud^n}{\ud\omega^n}\left[\frac{\exp(-i\omega\tau)}{(\omega-\Omega_b)^{n+1}}\right]
      \right\}.
\end{multline}
The computation of the derivatives is performed by using the formula
\begin{multline}
   \frac{\ud^n}{\ud\omega^n}\left[f(\omega)g(\omega)\right]=\sum_{p=0}^n\binom{n}{p}f^{(p)}(\omega)g^{(n-p)}(\omega)
\\
   \text{where}\quad
   \binom{n}{p}=\frac{n!}{p!(n-p)!}
\end{multline}
is the binomial coefficient. For the first term, $f(\omega)=(\omega+\Omega_b)^{-n-1}$ and
$g(\omega)=\exp(-i\omega\tau)$. This gives
\begin{multline}
   f^{(p)}(\omega)=(-n-1)(-n-2)\ldots(-n-p)(\omega+\Omega_b)^{-n-p-1}
\\
   =\frac{(-1)^p(n+p)!}{n!(\omega+\Omega_b)^{n+p+1}},
\end{multline}
and
\begin{equation}
   g^{(n-p)}(\omega)=(-i\tau)^{n-p}\exp(-i\omega\tau).
\end{equation}
A similar computation is done for the second term which finally leads to
\begin{multline}\label{g_n}
   \tilde{G}_n(\tau)=i\he(\tau)\sum_{p=0}^n\frac{(n+p)!}{n!p!(n-p)!}\frac{(i\tau)^{n-p}}{(2\Omega_b)^{n+p+1}}
\\\times
      \left[\exp(-i\Omega_b\tau)+(-1)^{n+p+1}\exp(i\Omega_b\tau)\right].
\end{multline}
A slight reorganization of the terms yields
\begin{multline}
   \tilde{G}_n(\tau)=\frac{\he(\tau)\tau^{n+1}}{(2\Omega_b)^nn!}\frac{1}{2\Omega_b\tau}
      \sum_{p=0}^n\frac{(n+p)!}{p!(n-p)!(2\Omega_b\tau)^p}
\\\times
      \left[i^{p-n-1}\exp(i\Omega_b\tau)+(-i)^{p-n-1}\exp(-i\Omega_b\tau)\right].
\end{multline}
We recognize the finite series expansion of the spherical Bessel function $j_n$~\cite{BABUSHKINA-1988} which finally gives
\begin{equation}
   \tilde{G}_n(\tau)=G_n(t,t')=\frac{\he(\tau)\tau^{n+1}}{(2\Omega_b)^nn!}j_n(\Omega_b\tau).
\end{equation}
This is the result given in the main text.

\section{Transfer matrix method}

The numerical simulations are based on a transfer matrix method. The idea is to link the wave amplitude just after the
$(j-1)$-th kick (\ie, $\psi_{j-1}$) to the wave amplitude just after the $j$-th kick (\ie, $\psi_j$). These fields are splitted
into two components: $\psi_j^+=\alpha_j\exp[-i\Omega_b(t-t_j)]$ propagating in the forward space direction and
$\psi_j^-=\beta_j\exp[i\Omega_b(t-t_j)]$ propagating in the backward direction. To obtain the expression of the transfer
matrix, we simply use the Lippmann-Schwinger expression of Eq.~(3) of the main text for the $j$-th kick. This gives for
$t>t_j$
\begin{multline}
   \alpha_je^{-i\Omega_b(t-t_j)}+\beta_je^{i\Omega_b(t-t_j)}
\\
      =\alpha_{j-1}e^{-i\Omega_b(t-t_{j-1})}+\beta_{j-1}e^{i\Omega_b(t-t_{j-1})}
\\
      +v_jG_b(t,t_j)\left[
         \alpha_{j-1}e^{-i\Omega_b(t_j-t_{j-1})}+\beta_{j-1}e^{i\Omega_b(t_j-t_{j-1})}
      \right].
\end{multline}
In matrix form, this gives by identification of the $\exp(i\Omega_bt)$ and $\exp(-i\Omega_bt)$ terms
\begin{equation}
   \begin{bmatrix}
      \mystruct\alpha_j \\ \mystruct\beta_j
   \end{bmatrix}
   =
   \begin{bmatrix}
      \left(1-\frac{v_j}{2i\Omega_b}\right)e^{-i\Omega_b\Delta t_j} & -\frac{v_j}{2i\Omega_b}e^{i\Omega_b\Delta t_j}
   \\
      \frac{v_j}{2i\Omega_b}e^{-i\Omega_b\Delta t_j} & \left(1+\frac{v_j}{2i\Omega_b}\right)e^{i\Omega_b\Delta t_j}
   \end{bmatrix}
   \begin{bmatrix}
      \mystruct\alpha_{j-1} \\ \mystruct\beta_{j-1}
   \end{bmatrix}
\end{equation}
where $\Delta t_j=t_j-t_{j-1}$.
This recurrence relation is used together with the initial condition $\alpha_0=-1/(2i\Omega_b)$ and
$\beta_0=1/(2i\Omega_b)$ (\ie, background Green's function $G_b$ at $t'=0$) to compute the wave amplitude at any time for a
Dirac delta source term in a given configuration of kicks. Then, an average over many disorder configurations is
performed. It is interesting to note that the transfer matrix is exactly the same as for a space disorder modelled by a series
of kicks. The difference lies in the fact that some input and output variables are interchanged, which makes it
possible to take into account the space boundary conditions instead of the initial conditions.